\documentclass{ws-procs975x65}
\usepackage{graphicx}
\usepackage{mathrsfs}

\def\beq{\begin{equation}}
\def\eeq{\end{equation}}

\newcommand{\bo}{\raise-1mm\hbox{\Large$\Box$}}

\newcommand{\f}[2]{\frac{#1}{#2}}

\newcommand{\la}{\langle}
\newcommand{\ra}{\rangle}
\newcommand{\w}{\omega}
\newcommand{\kp}{\kappa}
\newcommand{\be}{\begin{equation}}
\newcommand{\ee}{\end{equation}}
\newcommand{\bea}{\begin{eqnarray}}
\newcommand{\eea}{\end{eqnarray}}

\usepackage{hyperref}
\usepackage{url}
\usepackage{xcolor}
\usepackage{color}
\definecolor{amaranth}{rgb}{0.9, 0.17, 0.31}
\definecolor{purple(munsell)}{rgb}{0.62, 0.0, 0.77}
\definecolor{americanrose}{rgb}{1.0, 0.01, 0.24}
\definecolor{palatinateblue}{rgb}{0.15, 0.23, 0.89}
\definecolor{royalblue(web)}{rgb}{0.25, 0.41, 0.88}
\definecolor{hanpurple}{rgb}{0.32, 0.09, 0.98}
\definecolor{beaublue}{rgb}{0.74, 0.83, 0.9}
\definecolor{carminered}{rgb}{1.0, 0.0, 0.22}
\definecolor{brightpink}{rgb}{1.0, 0.0, 0.5}
\definecolor{vividviolet}{rgb}{0.62, 0.0, 1.0}
\hypersetup{ linktoc=all,
    colorlinks, linkcolor={palatinateblue},
    citecolor={brightpink}, urlcolor={amaranth}}

\newcommand{\changeurlcolor}[1]{\hypersetup{urlcolor=#1}}  

\begin{document}

\title{Unitary evaporation via modified Regge-Wheeler coordinate}
\author{Aizhan Myrzakul\footnote{aizhan.myrzakul@nu.edu.kz} and Michael R.R. Good\footnote{michael.good@nu.edu.kz}}

\address{Physics Department,\\
Nazarbayev University,\\
Astana, Kazakhstan\\
}


\begin{abstract}
Constructing an exact correspondence between a black hole model, formed from the most simple solution of Einstein's equations, and a particular moving mirror trajectory, we investigate a new model that preserves unitarity. The Bogoliubov coefficients in 1+1 dimensions are computed analytically. The key modification limits the origin of coordinates (moving mirror) to sub-light asymptotic speed. Effective continuity across the metric ensures that there is no information loss. The black hole emits thermal radiation and the total evaporation energy emitted is finite without backreaction, consistent with the conservation of energy. 
\end{abstract}


\bodymatter


\vspace{0.5cm}

\section{Introduction}\label{sec:metric} 



The Hawking effect \cite{Hawking:1974sw} has an exact correspondence\cite{Good:2016oey} with the moving mirror \cite{Davies:1976hi, Davies:1977yv}. The specific and analytically known accelerated boundary condition on the quantum field, $\psi$, located at the origin of coordinates, $r=0$, corresponding to the location of the black hole singularity, depends explicitly on the form of the tortoise coordinate, $r^*$. The moving mirror perfectly reflects the field modes and accelerates with the precise trajectory, $t(x)$. 


The physical effect of the mirror is that it arouses quantum field fluctuations reflecting virtual particles into real ones. The black mirror\cite{Good:2016oey}, which is the aforementioned tortoise coordinate associated boundary condition, extracts energy indefinitely and does not preserve unitarity. However, we present a summary of a modified model\cite{Good:2016atu} - ``a drifting black mirror" - which resolves these two problems. It was demonstrated recently \cite{GTU} that the new corresponding coordinate to $r^*$ is the generalized or giant tortoise coordinate (GTC). The giant tortoise coordinate results in unitarity preservation and finite energy emission of the black hole during evaporation.


In this MG15 proceedings contribution, first, the usual tortoise coordinate, $r^*$, and its relation to the moving mirror, $t(x)$, is briefly considered. Then we generalize this coordinate to the giant tortoise coordinate (GTC) and investigate the correspondence between the black hole and the moving mirror in the context of the GTC. We find no information loss, finite evaporation energy, thermal equilibrium, analytical beta coefficients, and a left-over\cite{Wilczek:1993jn} remnant.  



\section{The Tortoise Coordinate and the Black Mirror}
In this section the textbook matching solution\cite{Fabbri:2005mw} for the outside and inside of the black hole over the shock wave is derived. 

Let us start from the usual tortoise coordinate, (the Regge-Wheeler $r^*$):
\begin{equation}
r^*\equiv r+2M \ln\left(\frac{r}{2M}-1\right). \label{tortoise}
\end{equation}
Requiring the metric to be the same on both sides of the shock wave, $v_0$:
\begin{equation}
r(v_0, u_{\textrm{in}})=r(v_0, u_{\textrm{out}}), 
\end{equation}where
\begin{equation}
r(v_0, u_{\textrm{in}})=\frac{v_0-u_{\textrm{in}}}{2},\quad \quad 
\text{and} \quad\quad
r^*(v_0, u_{\textrm{out}})=\frac{v_0-u_{\textrm{out}}}{2},\label{4}
\end{equation} 
the tortoise coordinate, Eq.~(\ref{tortoise}), can be rewritten as: 
\be r(v_0,u_{\textrm{out}}) + 2M \ln\left(\frac{r(v_0,u_{\textrm{out}})}{2M}-1\right) = \frac{v_0 - u_{\textrm{out}}}{2}.\label{matched}\ee
Solving this for the red-shift function $u_\textrm{out}$ yields:
\begin{equation}
u_{\textrm{out}}=u_{\textrm{in}}-4M \ln\frac{|v_H-u_{\textrm{in}}|}{4M}.  \label{m1}
\end{equation}
The $u_{\textrm{out}}=+\infty$ limit corresponds to the formation of a black hole event horizon location, $v_H\equiv v_0-4M$. Eq.~(\ref{m1}) is exactly the matching solution\cite{Fabbri:2005mw} for the Eddington-Finkelstein background (exterior) to the Minkowski background (interior) with a strict event horizon.


Substituting $ u_{\textrm{out}} \equiv t(x) - x$ and $u_{\textrm{in}} \equiv t(x) + x$ into Eq.~(\ref{m1}), and solving for $t(x)$ gives the time-space trajectory of the black mirror\footnote{The black mirror, which is not eternally thermal\cite{Good:2012cp}, is called Omex for short, due to the \textbf{Om}ega constant, $\Omega e^{\Omega} = 1$, and \textbf{ex}ponent argument\cite{Good:2016oey}.}:
\be t(x) = v_H - x - 4M e^{x/2M},\label{omex}\ee
as investigated\footnote{See also the proceedings of the MG14 meeting\cite{Good:2015jwa,Anderson:2015iga} and the 2nd LeCosPA Symposium\cite{Good:2016bsq}.} in Good-Anderson-Evans (2016) \cite{Good:2016oey}.  
The range of the coordinates are: $0<r<\infty$ and $-\infty < x < \infty$. This is a transcendentally invertible and analytic relation between the black hole matching solution, Eq.~(\ref{m1}), and the trajectory of its moving mirror, Eq.~(\ref{omex}).

\section{The Giant Tortoise Coordinate and the Drifting Black Mirror}

We impose a strong restriction on the maximum speed of the black mirror: it must always travel slower than light, even asymptotically\footnote{An exception in a different model can give finite energy and preserve information if the acceleration asymptotes to zero sufficiently fast, see \cite{{Good:2016reflectatlight}}.}.  That is, in any coordinate system the origin of the black hole, $r=0$, should not speed away to null-future infinity, $\mathscr{I}^+$. 
This is highly restrictive and gives a new time-space trajectory \cite{Good:2016atu} of the origin of coordinates (an asymptotically drifting mirror):
\be t(x,\xi) = v_H - \frac{x}{\xi} - 4 M e^{\frac{x}{2M\xi}}. \label{domex}\ee
Here $\xi$ is the asymptotic drifting speed which lies in the range $0 < \xi < 1$. The corresponding matching condition can be easily derived and is:
\be u_{\textrm{out}} = u_{\textrm{in}} - 4 M \xi \ln\left[\frac{1-\xi}{2}\mathcal{W}\left(\frac{2 e^{\frac{ v_H - u_{\textrm{in}}}{2 M (1-\xi)}}}{1-\xi}\right)\right], \label{m2} \ee
where $\mathcal{W}$ is the product log. 
When $\xi \to 1$, the matching solution Eq.~(\ref{m2}) is equivalent to Eq.~(\ref{m1}), 
meaning that one has operative formation of an effective event horizon.  Thus the modest modification in Eq~(\ref{domex}) safeguards the formation of a black hole \footnote{For more information on whether any type of horizon is formed during gravitational collapse taking into account quantum effects see, e.g.\cite{Mann:2018jcf}. For horizonless models see\cite{Good:2017kjr, Good:2017ddq}.} that occurs with Eq~(\ref{omex}).

Eq.~(\ref{m2}) represents the world line of the origin.  It is easy to see that this origin is a perfectly reflecting boundary as nothing can go behind $r=0$ into the $r<0$ space.   
Therefore we use the generalized matching solution Eq.~(\ref{m2}) in order to find the giant tortoise coordinate (analogous to going backwards from Eq.~(\ref{m1}) to Eq.~(\ref{tortoise})),
\be \label{giant} r^*(\xi) \equiv r+ 2 M \xi  \ln \left[\frac{1-\xi}{2} \mathcal{W}\left(\frac{2 e^{\frac{r-2 M}{M (1-\xi)}}}{1-\xi}\right)\right].\ee
It is a crucial fact that these two coordinates, $r^*$ and $r^*(\xi)$, are effectively the same when $\xi \approx 1$.  The distinction is that there is no singularity at $r = 2M$ in Eq.~(\ref{giant}) as in Eq.~(\ref{tortoise}) when $\xi \neq 1$: $r^*(\xi)_{r=2M} = 2M \left[1- \xi \mathcal{W}(2/\epsilon)\right]$, where $\epsilon \equiv 1-\xi$.  The user may define the free parameter $\xi$ as close to $\xi \approx 1$ (for effective continuity) as is arbitrarily desired as long as strictly, $\xi < 1$ (for unitarity).  \\
 
\section{Unitarity: Finite Asymptotic Entanglement Entropy }

Qualitatively, the black mirror correspondence demonstrates information loss by the acceleration horizon which prohibits some left-movers from hitting the mirror and becoming right-movers.  We can say these modes are lost forever in the black hole.  However, the drifting black mirror has an asymptotic approach to time-like future infinity, $i^+$, rather than null future infinity, $\mathscr{I}^+_L$ and all the left-movers hit the mirror and become right-movers, preserving information to an observer at $\mathscr{I}^+_R$.  

Quantitatively, we see this result by the von-Neumann entanglement entropy\cite{Chen:2017lum} for the black mirror\cite{Good:2016atu}: 
\be S(t) = \frac{1}{6} \textrm{tanh}^{-1}\left(\frac{1}{\mathcal{W}(2e^{-2\kappa t})+1}\right), \ee
whose limit in the far future diverges:
\be S_f \equiv \lim_{t\to\infty} S(t) = \infty, \ee
signaling information loss (see e.g. the entropy divergence in \cite{GoodS}). Here $\kappa \equiv 1/4M$, the surface gravity for Schwarzschild background.  In contrast, the drifting black mirror has entropy,
\be S_{\xi}(t) = \frac{1}{6} \textrm{tanh}^{-1}\left(\frac{\xi}{\mathcal{W}(2e^{-2\kappa t})+1}\right), \ee
whose limit is
\be S_f \equiv \lim_{t\to\infty} S_{\xi}(t) = \frac{1}{6}\textrm{tanh}^{-1}(\xi) = \frac{\eta}{6} \neq \infty. \ee
The final asymptotic entropy is the drifting rapidity and its measure as a finite constant signals information preservation. To achieve effective equilibrium ($\xi \approx 1$), the final asymptotic entropy will be very large ($\eta \ggg 1$), but finite.
\section{Finite Evaporation Energy}
A prime advantage of the GTC, Eq.~(\ref{giant}), is that during the collapse the global energy emission of the black hole is finite and analytic. The consistency of the result with the analytically computed beta Bogolubov coefficient can be shown via a numerical verification of the stress-energy tensor \cite{Good:2016atu, Good:2013lca} whose total energy production is,
\be E = \frac{1}{96\pi M}\left(\gamma^2 + \frac{\eta}{\xi}\right),
\ee
where $\gamma \equiv 1/\sqrt{1-\xi^2}$ is the final drifting Lorentz factor, $\eta \equiv \tanh^{-1}\xi$ is the final drifting rapidity, with $\xi<1$ corresponding to the final drifting speed.  
For high drifting speed (thermality), $\xi \approx 1$, then $\gamma^2 \gg \eta/\xi$, and:
\be E=\frac{\gamma^2}{96\pi M}. \label{energy} \ee 
One immediately sees that the energy diverges as the origin moves to the speed of light, (i.e. mirror drift, $\xi \to 1$ and $\gamma \to \infty$). Eq.~(\ref{energy}) is the final expression for the energy emission of the thermal black hole which is finite and consistent with the conservation of energy.  Finite energy is anticipated to be a primary result of backreaction, yet, we have obtained consistency by restricting origin speed, $\xi < 1$.
\section{Temperature and the Giant Tortoise Coordinate }
Using the GTC, a constant energy flux plateau is apparent (high drifting speeds, $\xi \approx 1$).  During equilibrium, $F = \pi T^2/12$. Expanding the temperature as a function of maximum energy flux (where the radiation is closest to equilibrium) gives the temperature of the black hole  \cite{Good:2016atu}:
\begin{equation}
T(\epsilon,M) = \frac{1}{8\pi M}\left[1-3\left(\frac{3}{4}\right)^{1/3}\epsilon^{2/3}+O\left(\epsilon\right)\right]\label{hot},
\end{equation} to lowest order in  $\epsilon$ where $\epsilon \equiv 1-\xi$. The first term in this expression corresponds to the equilibrium temperature of the unmodified black mirror model which uses the usual tortoise coordinate.  The other terms correspond to the deviation due to sub-light speed drift, $\xi < 1$, which are negligible for small $\epsilon$. The modification of Eq.~(\ref{omex}) to Eq.~(\ref{domex}) still results in a constant energy flux plateau and effective long-term thermal equilibrium of Eq.~(\ref{hot}) for $\xi \approx 1$. This confirms the robustness of the GTC model for describing Planckian distributed particle creation from a black hole.
\section{The Beta Bogoliubov Coefficient}

The particle spectrum,
\begin{equation}
\left\langle N_\omega\right\rangle\equiv\langle 0_{in}|N_\omega^{out}|0_{in}\rangle=\int_0^\infty|\beta_{\omega\omega^{'}}|^2d\omega^{'},\label{spectrum}
\end{equation}
requires knowing the GTC beta Bogoliubov coefficient which is a simple integral to compute analytically\cite{Good:2016atu} with result:
\begin{equation}
\beta_{\omega\omega'}(\xi) = -\frac{\xi\sqrt{\omega\omega'}}{2\pi\kappa \omega_p} \left(\frac{i\kappa}{\omega_p}\right)^{A} \Gamma(A),\label{beta}
\end{equation}
where $\Gamma(x)$ is the gamma function, $A \equiv \frac{i}{2\kappa}[(1+\xi)\omega + (1-\xi)\omega']$ and $\omega_p \equiv \omega+\omega'$. The integrand\cite{Good:2016atu} of Eq.~(\ref{spectrum}) is Planckian using Eq.~(\ref{beta}) with $\xi \approx 1$ and $\omega'\gg \omega$, which is consistent with Eq.~(\ref{hot}).  \\

\section{Summary}

\begin{tabular}{|c|c|c|}
	\hline
	Quantity & Tortoise & Giant Tortoise\\
    \hline
    $r^*$ & $r^*\equiv r+2M \ln\left(\frac{r}{2M}-1\right)$ & $r^*(\xi)\equiv r+ 2 M \xi  \ln \left[\frac{\epsilon}{2} \mathcal{W}\left(\frac{2}{\epsilon} e^{\frac{r-2 M}{M \epsilon}}\right)\right]$\\
    \hline
    $t(x)$ & $t = - x - 4M e^{x/2M}$ & $t(\xi) = - x/\xi - 4 M e^{x/2M\xi}$\\
    \hline
    $u_\textrm{out}$ & $u_{\textrm{out}}=u_{\textrm{in}}-4M \ln\frac{|v_H-u_{\textrm{in}}|}{4M}$ & $u_{\textrm{out}}(\xi) = u_{\textrm{in}} - 4 M \xi \ln\left[\frac{\epsilon}{2}\mathcal{W}\left(\frac{2}{\epsilon} e^{\frac{ v_H - u_{\textrm{in}}}{2 M \epsilon}}\right)\right]$ \\ 
    \hline
    $S_f$ & $\infty$ & $S_f(\xi) = \eta/6$\\
    \hline 
    $E$ & $\infty$ & $E(\xi) = \frac{1}{96\pi M}\left(\gamma^2 + \frac{\eta}{\xi}\right)$\\
    \hline
    $T$ & $T = \frac{1}{8\pi M}$ & $T(\xi) = \frac{1}{8\pi M} + \mathcal{O}(1-\xi),~~\textrm{for} ~~\xi\approx 1$\\
    \hline
    $\beta$ & $\beta_{\omega\omega'} = -\frac{\sqrt{\omega\omega'}}{2\pi\kappa \omega_p} \left(\frac{i\kappa}{\omega_p}\right)^{\frac{i\omega}{\kappa}} \Gamma(\frac{i\omega}{\kappa})$ & $\beta_{\omega\omega'}(\xi)=-\frac{\xi\sqrt{\omega\omega'}}{2\pi\kappa \omega_p} \left(\frac{i\kappa}{\omega_p}\right)^{A} \Gamma(A)$\\
    \hline
   
\end{tabular}
\\\\

We have presented an overview of a unitary black hole evaporation model that, without backreaction, manages to produce a finite total energy emission.  Exact analytic results for several important corresponding quantities are found: the beta Bogoliubov coefficients, the finite total energy emission, the matching condition for the modes and the generalized tortoise coordinate (GTC).  

The model\footnote{The analytically solvable theory exploits the special dual properties of conformal flatness/invariance.\cite{Fulling:2018lez}} relaxes uncompromising continuity across the shock wave in the metric in exchange for preserving information. Arbitrary precision in continuity is permitted with arbitrary fast sub-light drifting speeds for the origin of coordinates (the moving mirror must travel at speeds less than light).  As long as this requirement is met, the information is preserved and the energy emitted is finite.  With ultra-relativistic, sub-light speeds the black hole emits particles in a Planck distribution with constant energy flux at equilibrium temperature.  \\


\end{document}